\documentclass[aps,prl,superscriptaddress,twocolumn,showpacs,amsmath,floatfix,citeautoscript]{revtex4}

\usepackage{graphicx}
\usepackage{float}
\usepackage{dcolumn}
\usepackage{color}
\usepackage{latexsym,bm}
\usepackage[normalem]{ulem}
\usepackage{multirow}

\begin{document}

\hyphenpenalty=5000

\tolerance=1000

\title{Gapless spin liquid ground state of spin-1/2 $J_1$-$J_2$  Heisenberg  model on square lattices}

\author{Wen-Yuan Liu}
\affiliation{Key Laboratory of Quantum Information, University of Science and
  Technology of China, Hefei, Anhui, 230026, People's Republic of China}
\affiliation{Synergetic Innovation Center of Quantum Information and Quantum
  Physics, University of Science and Technology of China, Hefei, 230026, China}
%  \affiliation{Department of Physics, The Chinese University of Hong Kong, Shatin, New Territories, Hong Kong, China}

\author{Shaojun Dong}
\affiliation{Key Laboratory of Quantum Information, University of Science and
  Technology of China, Hefei, Anhui, 230026, People's Republic of China}
\affiliation{Synergetic Innovation Center of Quantum Information and Quantum
  Physics, University of Science and Technology of China, Hefei, 230026, China}

\author{Chao Wang}
\affiliation{Key Laboratory of Quantum Information, University of Science and
  Technology of China, Hefei, Anhui, 230026, People's Republic of China}
\affiliation{Synergetic Innovation Center of Quantum Information and Quantum
  Physics, University of Science and Technology of China, Hefei, 230026, China}

\author{Yongjian Han}
\email{smhan@ustc.edu.cn}
\affiliation{Key Laboratory of Quantum Information, University of Science and
  Technology of China, Hefei, Anhui, 230026,  People's Republic of China}
\affiliation{Synergetic Innovation Center of Quantum Information and Quantum
  Physics, University of Science and Technology of China, Hefei, 230026, China}

\author{Hong An}
\affiliation{School of Computer Science and Technology,University of Science and Technology of China, Hefei, 230026,
  China}

\author{Guang-Can Guo}
\affiliation{Key Laboratory of Quantum Information, University of Science and
  Technology of China, Hefei, Anhui, 230026,  People's Republic of China}
\affiliation{Synergetic Innovation Center of Quantum Information and Quantum
  Physics, University of Science and Technology of China, Hefei, 230026,
  China}

\author{Lixin He}
\email{helx@ustc.edu.cn}
\affiliation{Key Laboratory of Quantum Information, University of Science and
  Technology of China, Hefei, Anhui, 230026,  People's Republic of China}
\affiliation{Synergetic Innovation Center of Quantum Information and Quantum
  Physics, University of Science and Technology of China, Hefei, 230026, China}
\date{\today }


\begin{abstract}
The spin-1/2 $J_1$-$J_2$ Heisenberg model on square lattices are investigated via
the finite projected entangled pair states (PEPS) method.
Using the recently developed gradient optimization method combining with Monte Carlo sampling techniques,
we are able to obtain the ground states energies that are competitive to the best results.
The calculations show that there is no N\'eel order, dimer order and plaquette order
in the region of 0.42 $\lesssim J_2/J_1\lesssim$ 0.6, suggesting a single spin liquid phase in the intermediate region. Furthermore, the calculated staggered spin, dimer and plaquette correlation functions all have power law decay behaviours, which provide strong evidences that the intermediate nonmagnetic phase is a single gapless spin liquid state.

\end{abstract}
\maketitle

\date{\today}

%\section{Introduction}

During the past decades, the frustrated magnets have attracted enormous attentions~\cite{frustrated}.
The frustrated interactions result in a large degeneracy of the ground state,
and the quantum fluctuation may lead to massive coherent superposition of the
degenerated states, implying a novel highly entangled (correlated)
quantum state, known as quantum spin liquid (QSL)~\cite{Anderson1973,Anderson1987,Balents2010},
which lacks any long range magnetic order even down to zero temperature.
Because of the anomalously high degree of entanglement, QSLs have nontrivial topological properties
which may host exotic excitations with fractional statistics, such as spinons, and visions, etc.,
which have important applications in quantum computing\cite{Kitaev2003,Kitaev2006}.

The spin-1/2 $J_1$-$J_2$ Heisenberg model on  square lattices is one of the primary candidate models to study the QSL, which was first introduced to describe the breakdown of N\'eel antiferromangetic (NAF) long-range order (LRO) in cuprate superconductors\cite{inui1988,chandra1988,dagotto1989}.
It is widely accepted that this model exhibits an NAF LRO at small $J_2/J_1$ region and a collinear antiferromangetic LRO at large $J_2/J_1$, separated by a nonmagnetic phase in the region of $0.4\lesssim J_2/J_1 \lesssim 0.6$.

Despite extensive investigation in the past three decades by various methods\cite{dagotto1989,figueirido1990,schulz1992,schulz1995,
schulz1996,chandra1988,ivanov1992,sirker2006,beach2009,murg2009,mambrini2006,
capriotti2000,capriotti2001,schmalfu2006,mezzacapo2012,darradi2008,jiang2012,hu2013,gong2014,
chou2014,satoshi2015,richter2015,wang2016,poilblanc2017,wang2017,haghshenas2017}, the nature of the nonmagnetic region is
still highly controversial.
Early density matrix renormalization group (DMRG) study\cite{jiang2012} suggested that the nonmagnetic region is a gapped $Z_2$ spin liquid phase. However, more recently DMRG study with SU(2) symmetry \cite{gong2014} suggested a plaquette valance bond (PVB) phase for 0.5$\lesssim J_2/J_1\lesssim $0.61 with a near critical region for  $0.45\lesssim J_2/J_1\lesssim 0.5$.
On the other hand, variational quantum Monte Carlo (vQMC) simulations\cite{hu2013} suggested the nonmagnetic region is a gapless QSL.
Therefore, the understanding of the true nature of the nonmagnetic region is still far from complete.

 Recently developed  projected entangled pair states (PEPS) method\cite{verstraete2004}, provides a new powerful tool to simulate two-dimensional quantum many-body systems. Unlike the DMRG method, the PEPS satisfies area law in two dimensions, and therefore is a more natural way to study the strongly correlated systems in two dimensions. However, PEPS methods suffer from extremely high computational scaling to the virtual bond dimensions, and are very difficult to optimize.
 %previous PEPS calculations give much higher Neel to nonmagnetic transition $J_2$ \cite{??}(\red{\it Please check, and cite relavant refs. including Sheng DN's paper}) compared to other methods.
 Recently, we developed a finite PEPS optimization algorithm which combines the stochastic gradient optimization and Monte Carlo (MC) sampling techniques\cite{liu2017,he2018}. It can give much higher precision than the simple update\cite{simpleupdate} and even full update methods\cite{fullupdate2014}, making it a reliable method to investigate the properties of the intermediate nonmagnetic phase.

In this paper, we investigate the ground state of the nonmagnetic phases of the $J_1$-$J_2$ model using our recently
developed finite PEPS methods.
We find that for 0.42$\lesssim J_2/J_1\lesssim $0.6, the spin order, as well as the dimer order all vanish in the thermodynamic limit, which rules out the possibility of valence-bond  solid (VBS) states including PVB and columnar valance bond (CVB) \cite{sachdev1990,chubukov1991,singh1990,haghshenas2017}, and no additional phase transitions are found in this region. Furthermore, both the calculated spin-spin and dimer-dimer correlations show power law decay, suggesting that the region is a gapless QSL. These results are consistent with the recent
vQMC simulations\cite{hu2013}.

%\section{The Model and Methods}
The spin-1/2 $J_1$-$J_2$ Heisenberg model is given by,
\begin{equation}
H=J_1\sum_{\langle i,j\rangle}{\bf S}_i \cdot {\bf S}_j+J_2\sum_{\langle\langle i,j\rangle\rangle}{\bf S}_i \cdot {\bf S}_j  ~~,
\end{equation}
where $\langle i,j\rangle$ and $\langle\langle i,j\rangle\rangle$ denote the nearest-neighbor (NN) and the next-nearest-neighbor (NNN) spin pairs respectively. We assume the exchange couplings $J_1$,$J_2$ $>$0. Without loss of generality, we set $J_1$=1 throughout this paper.

We study the system on an $L$$\times$$L$ square lattice with open boundary conditions, for $L$ up to 16.
We represent the ground state wave functions by PEPS with virtual bond dimension $D$. All parameters in the PEPS wave functions are independent, and subject to optimization. When optimizing the PEPS, we first perform imaginary time evolution with the simple update method\cite{simpleupdate}.
We then further optimize the PEPS using the stochastic gradient method until the results are fully converged\cite{liu2017}.
The energies, and energy gradients are calculated via the MC sampling technique.
The method greatly improves the ground state energies compared to the simple update and even full update method\cite{fullupdate2014}.
More details about the method can be found in Ref.~\onlinecite{liu2017}. With sufficiently optimized
ground state PEPS wave functions, the physical quantities and correlation functions, including
the staggered magnetization, dimer and plaquette order parameters, spin-spin correlations, dimer-dimer correlations, and plaquette-plaquette correlations are calculated via Monte Carlo sampling techniques.
% with 500000 sweeps.

To guarantee the reliability of the calculations, the convergence to the virtual dimension $D$ and the truncation dimension $D_c$ during contractions are carefully checked.
We find that $D=8$, $D_c=24$ are enough for the systems up to 16$\times$16
(See the Supplemental Materials). All results are obtained under these parameters, unless otherwise claimed.

%\section{Ground State Energies}
\begin{figure}
 \centering
 \includegraphics[width=3.2in]{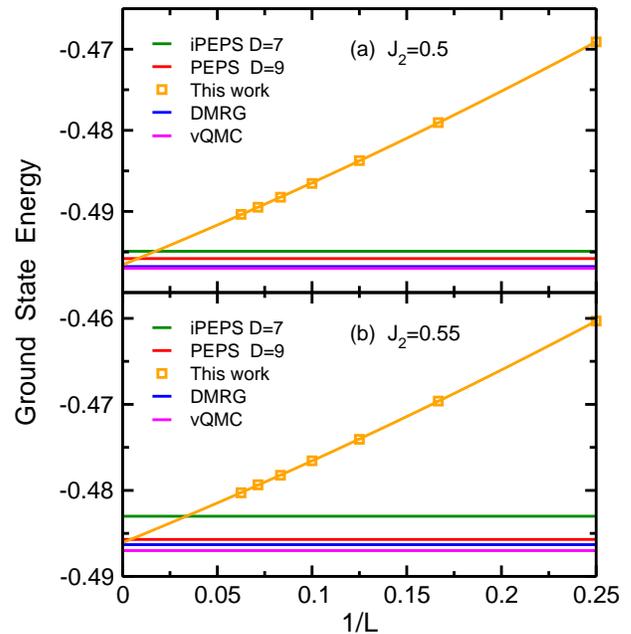}
 \caption{Ground state energies in the 2D limit for (a) $J_2$=0.5 and (b) $J_2$=0.55 obtained by second order polynomial extrapolations of the energies at $L=4-16$. The horizontal straight lines denote the ground state energies from
 previous calculations in the literatures, where the green, red, blue, magenta lines are the ground state energies in 2D limit obtained by iPEPS with $D$=7~\cite{poilblanc2017}, finite PEPS $D$=9 based on periodic systems\cite{wang2016}, DMRG with SU(2) symmetry\cite{gong2014}, and the variational quantum Monte Carlo plus Lanczos extrapolation\cite{hu2013}.
  The values of the ground state energies in the 2D limit are listed in the Supplemental Materials Table S3.}
 \label{fig:Energy_J05andJ055}
 \end{figure}

The ground state energies at different $J_2$, particularly in  the highly frustrated region, are important criterions for the precision of a computational method. We calculate ground state energies of different $J_2$ for system size $L$=4$-$16. We then perform finite size scaling to obtain the ground state energies in the thermodynamic limit.
In our previous studies\cite{liu2017}, it has been shown for Heisenberg model, i.e., $J_2$=0,
 the ground state energy per site obtained by $D$=10 is $E_0=-0.66948(42)$, in excellent agreement with the quantum Monte Carlo result $E_0=-0.669437(5)$\cite{MCHeisenberg1997}.

We show the ground state energies for $J_2$=0.5 and 0.55 with different system sizes in  Fig.~\ref{fig:Energy_J05andJ055}(a),(b) respectively. The extrapolated energies at the thermodynamic limit are $E_1=-0.4966(1)$ for $J_2=0.5$ and $E_2=-0.4861(1)$ for  $J_2=0.55$.
Some previous calculated ground state energies in the literatures are also shown for comparisons.
The ground state energies obtained in this work are significantly lower than previous iPEPS results with $D$=7 \cite{poilblanc2017} and the finite PEPS calculation with $D$=9\cite{wang2016}. These energies are almost the same as the best DMRG results\cite{gong2014}  $E_1\simeq -0.4968$ and $E_2\simeq -0.4863$, obtained by a rough estimation based on cylindrical geometries; they are also comparable to the energies from vQMC plus Lanczos extrapolation~\cite{hu2013}.

%{\it We would like to point out that the above energies are only based on $D$=8 (we do not resort to the extrapolating of $D$). The extrapolating energyies based on bond dimension $D$ ranging from 3 to 8 and system size $L$=4$-$16 for $J_2$=0.5 and 0.55 can be found in the Supplementary Material.
%\red{\it Do we want say the convergency of energies to D?}
%
%
%%\section{Order Parameters}

\begin{figure}
 \centering
 \includegraphics[width=3.2in]{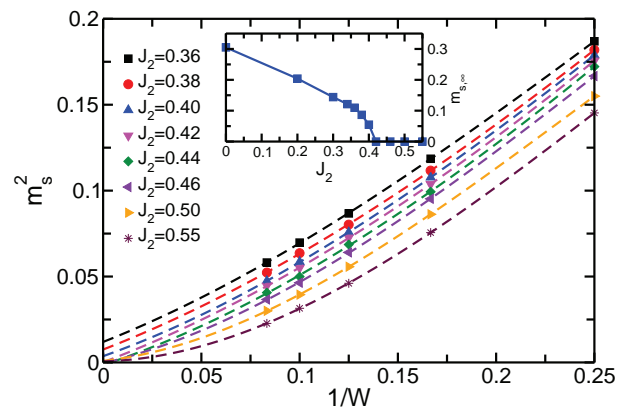}
 \caption{The N\'eel order parameter $m_s^2$ calculated by PEPS with $D$=8, on the $L$=8,10,12,14,16 square lattices where the central bulk size is $W$=$L-$4. Extrapolations to 2D limit are performed with a third order polynomial fitting.
 The inset depicts the $m_s$ in the thermodynamic limit at different $J_2$.  }
 \label{fig:spinStructFactor}
 \end{figure}

With the fully optimized ground states, we investigate the nature of the ground state. We first measure the N\'eel order parameter $m^2_s=\frac{1}{N^2}\sum_{ij}{\langle{\bf S}_i \cdot {\bf S}_j}\rangle {e}^{i {\bf k}\cdot({\bf r}_i-{\bf r}_j)}$ with $\bf k=(\pi,\pi)$ to distinguish the magnetic and the nonmagnetic phases at different $J_2$.
 To minimize the boundary effects, the summations are restricted in the central $W$$\times$$W$ lattice~\cite{liu2017}, and here $W$=$L-4$ are used.
At $J_2$=0, the calculated staggered magnetization is $m_{s,\infty}$=0.305~\cite{liu2017}, which is
in excellent agreement with the QMC result $m_{s,\infty}$=0.307\cite{MCHeisenberg1997}. We present $m_s^2$ for different system sizes with $L$=8$-$16 in the Fig.~\ref{fig:spinStructFactor}, and $m_s$ for the 2D limit in the inset of Fig.\ref{fig:spinStructFactor}. These results suggest that the magnetic to nonmagnetic phase transition
is located at $J_2\simeq$ 0.42, falling in the range of previous studies
 0.41 $\sim$ 0.45~\cite{jiang2012,hu2013,gong2014,chou2014,satoshi2015,richter2015}.
%\red{We note that previous PEPS calculations overestimate the transition point $J_2$.}

% and several states  have been proposed as the candidate states, such as plaquette valence-bond (PVB) states\cite{zhitomirsky1996,capriotti2000,takano2003,mambrini2006,isaev2009,yu2012,doretto2014,gong2014} , columnar valence-bond (CVB) states\cite{sachdev1990,chubukov1991,singh1990,haghshenas2017}, a gapless QSL\cite{capriotti2001,zhang2003,wang2013,hu2013}, and a gapped $Z_2$ QSL\cite{jiang2012}.
%Previous DMRG calculations claim that the intermediate phase is a Z2 gapped QSL.\cite{??}
%Recent DMRG calculations suggest that there are actually two phases in this region: for $J_2$ between ?? -??, there is a gapless region, and for $J_2$ between ?? to ?? there is a PVB state.}
%
%\red{\it Do we want to compare the results to simple update's results?}

However, the exact nature of the intermediate nonmagnetic phase is still under intensive debates.
We need further clarify the nature of the nonmagnetic region,
especially, to answer: whether the phase is a QSL phase or a VBS phase ? Is there a phase transition from QSL to VBS ?
We calculate the dimer structure factors which can be used to detect the possible VBS order,
\begin{equation}
M^{\alpha}_d({\bf k})=\frac{1}{N}\sum_{ijkl}(\langle B^{\alpha}_{i,j}  B^{\alpha}_{k,l}\rangle-\langle B^{\alpha}_{i,j}\rangle \langle B^{\alpha}_{k,l}\rangle) {e}^{i {\bf k}\cdot({\bf r}_i-{\bf r}_j)},
\end{equation}
where $\alpha=x,y$, and $B^{x}_{i,j}={\bf S}_{i,j} \cdot {\bf S}_{i+1,j}$ and $B^{y}_{i,j}={\bf S}_{i,j} \cdot {\bf S}_{i,j+1}$ are horizontal and  vertical bond operators along the $x$-axis and $y$-axis, respectively. The summation is restricted in the central bulk  $W$=$L-6$ to reduce boundary effects. The VBS order is indicated by peaks appearing at ${\bf k}$=$(\pi,0)$ for $M^{x}_d({\bf k})$ or at ${\bf k}$=$(0,\pi)$ for $M^{y}_d({\bf k})$. Therefore, one may define the horizontal and vertical dimer order parameters as  $m^{2}_{dx}$=$\frac{1}{N} M^{x}_d({\bf k})$ with  ${\bf k}$=$(\pi,0)$  and  $m^{2}_{dy}$=$\frac{1}{N} M^{y}_d({\bf k})$ with  ${\bf k}$=$(0,\pi)$, respectively.

\begin{figure}[tb]
 \centering
 \includegraphics[width=2.8in]{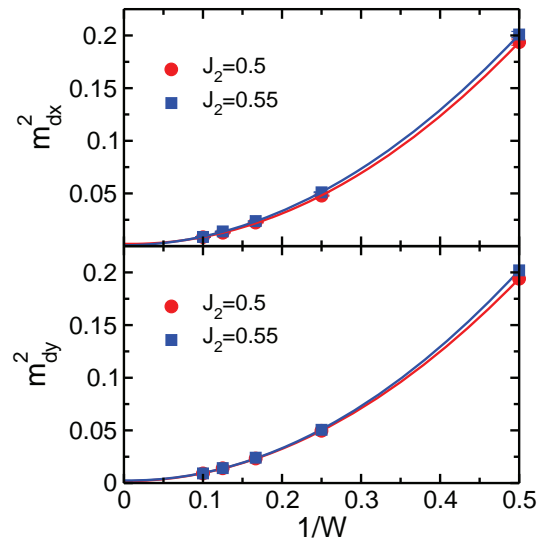}
 \caption{The horizontal and vertical dimer order parameters $m^2_{dx}$ and $m^2_{dy}$ for $J_2=0.5$ and 0.55 with system sizes $L=8-16$. Extrapolations are performed using second order polynomial fittings.}
 \label{fig:dimerStructFactor}
 \end{figure}

Figure~\ref{fig:dimerStructFactor} depicts the dimer order parameters $m^2_{dx}$ and $m^2_{dy}$,
 calculated at two typical nonmagnetic points, $J_2$=0.5 and 0.55 with different system sizes. % \red{Interestingly,  the dimer orders at $J_2$=0.5 and $J_2$=0.55 are very close to each other, and  for both $J_2$, the dimer orders decrease dramatically with increasing $L$, and vanish in the 2D limit.}
  According to the deconfined quantum critical point (DQCP) theory\cite{senthil2004}, the complex order parameter $m_{dx}+im_{dy}$ is sufficient to detect and distinguish both columnar and plaquette VBS phases. We find that $m^2_{dx}$ and $m^2_{dy}$ are almost the same within numerical precision at each lattice size, reflecting the isotropy of horizontal and vertical directions, which is expected for the true ground states and exclude the CVB phases. As a result, the vanishes of dimer orders in both $x$ and $y$ directions do not support a VBS order at $J_2$=0.5 and 0.55 at the thermodynamic limit, which indicate that the whole intermediate nonmagnetic region is actually a QSL phase and there is no phase transition to VBS phase which is different from the results of Gong et al.\cite{gong2014}.

%\red{\it Compare our results to Gong's and Wang L's results. Compare our results to VQMC results.}

% \section{ Correlation functions}

%There are different type of QSL are proposed: DMRG suggested a gapped $Z_2$ QSL\cite{jiang2012}, however, vQMC suggested a gapless QSL\cite{hu2013}.

To further explore the properties of the QSL, more explicitly, whether it is gapped or gapless, we calculate the staggered spin-spin, dimer-dimer and plaquette-plaquette correlation functions along straight lines.

\begin{figure}
 \centering
 \includegraphics[width=3.2in]{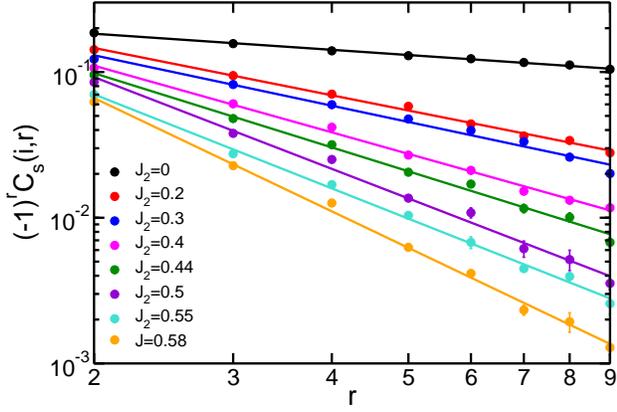}
 \caption{Log-log plots of spin-spin correlation functions versus distance on a 14$\times$14 lattice for different $J_2$.
 %Solid lines with  power law fittings are shown as a guide to the eye.
 }
 \label{fig:lineSpinSpin}
 \end{figure}

The spin-spin correlation functions are calculated on a 14$\times$14 lattice,
and the results are averaged over the central $M$=6 rows, %\red{\it You used $S^z$ in the following equation.
%Should you explain, or just change to ${\bf S}$?}
%
 \begin{equation}
 C_{s}(i,r)=\frac{1}{M}\sum_{j} \langle {\bf S}_{i,j}\cdot{\bf S}_{i+r,j} \rangle,
 \end{equation}
where $j$ is restricted in the central $M$ rows and $i$ is fixed to 2.
As shown in Fig.~\ref{fig:lineSpinSpin}, the spin-spin correlations have power law decay in a large parameter region, from $J_2$=0 to $J_2$=0.58. In the N\'eel phase a long range order will exhibit, and spin correlations are expected to eventually decay to a saturation value  theoretically. Due to our current computational limit, we can not access larger systems to observe such a saturation value, but it is notable that at $J_2$=0 the absolute valule of $C_{s}(i,r$=$9)\simeq$ 0.102 on 14$\times$14 lattice is very close to the  QMC value $C_{s}(i,r$$\rightarrow$$\infty)$$\simeq$ 0.094 on infinite system\cite{MCHeisenberg1997}, indicating the calculated power law decay behavior of spin-spin correlations obtained from finite systems for the N\'eel phase ($J_2 \lesssim 0.42$) is reliable to some extent.  The power law decay behaviors in the QSL phase imply that there is no ${\bf S}$=1 gap in 2D limit. The spin-spin correlation behaviors are consistent with the lack of VBS orders in the intermediate phase, in which the ${\bf S}$=1 gap is expected.
The power law decay exponents for different $J_2$ are listed in Table S4 of the Supplemental Materials,
which increase with the increasing of $J_2$.
We note that the decay exponents fitted from the finite systems here should not be compared directly to those of the infinite systems.

We further calculate the dimer-dimer and plaquette-plaquette correlation functions. The horizontal dimer-dimer correlations are defined as
 \begin{equation}
 C^{h}_{dx}(i,r)=\frac{1}{M}\sum_{j}(\langle B^{x}_{i,j} B^{x}_{i+r,j} \rangle-\langle B^{x}_{i,j}\rangle \langle B^{x}_{i+r,j} \rangle) \, .
 \end{equation}
Similarly, we can define the vertical dimer-dimer correlations.
% \begin{equation}
%  C^{v}_{dy}(j,r)=\frac{1}{M}\sum_{i}(\langle B^{y}_{i,j} B^{y}_{i,j+r} \rangle-\langle B^{y}_{i,j}\rangle \langle B^{y}_{i,j+r} \rangle).
% \end{equation}
The plaquette-plaquette correlations are defined as
  \begin{equation}
 C_{p}(i,r)=\frac{1}{M}\sum_{j}(\langle Q_{i,j}Q_{i+r,j} \rangle-\langle Q_{i,j}\rangle \langle Q_{i+r,j} \rangle),
 \end{equation}
 where $Q_{i,j}=\frac{1}{2}(P_{\Box,i,j}+P^{-1}_{\Box,i,j})$ and $P_{\Box,i,j}$ denotes the cyclic exchange operator of the four spins on a given plaquette.
 All correlation functions are averaged in the central $M$=4 odd rows on 16$\times$16 squares lattice and $i=3$.

\begin{figure}
 \centering
 \includegraphics[width=2.8in]{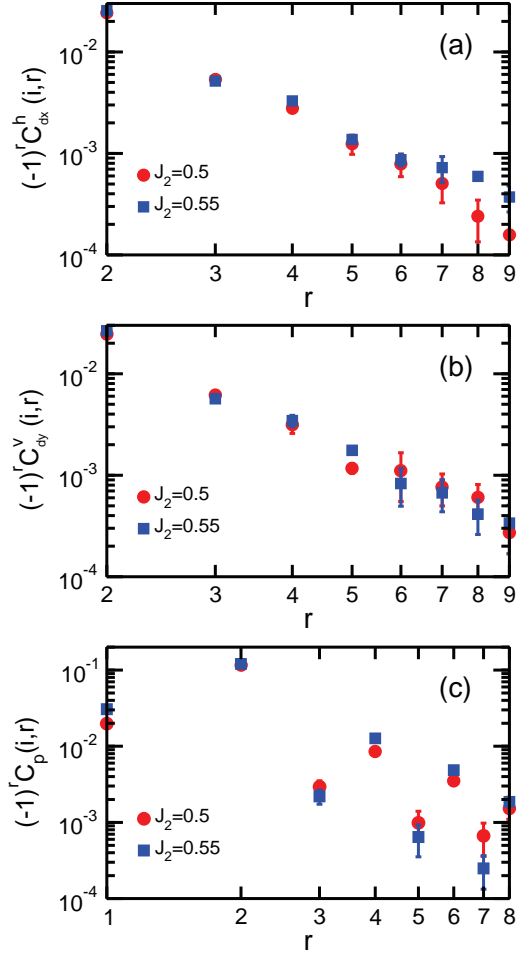}
 \caption{Log-log plots of (a) the horizontal dimer-dimer correlation functions along the $x$-axis, (b) the vertical dimer-dimer correlation functions along the $y$-axis, and (c) the plaquette-plaquette correlation functions along the $x$-axis on a 16$\times$16 lattice at $J_2$=0.5 and 0.55.}
 \label{fig:lineDimerDimer}
 \end{figure}

Figure~\ref{fig:lineDimerDimer}(a),(b) and (c) depict the staggered horizontal, vertical dimer-dimer correlations, and plaquette-plaquette correlations respectively at two typical points $J_2$=0.5 and 0.55. We find that both the dimer-dimer and plaquette-plaquette correlation functions have power law decay, indicating that there is no spin ${\bf S}$=0 gap.
The fitted power law decay exponents are about 2.8 for the dimer correlations.
The plaquette correlation functions show large oscillations for odd and even $r$, which maybe has close relation with the local  plaquette order existing in the finite system. The fitted power law decay exponents are about 1.8 -2.0 if only odd sites are used, and about 3.0 if only even sites are used in the fit.

 %\section{discussion and Conclusion}

The above results give strong evidences that the intermediate nonmagnetic phase is a gapless QSL, for there are no columnar orders or plaquette orders, and all correlation functions including spin-spin, dimer-dimer, plaquette-plaquette correlations have power law decay. These results are consistent with the conclusions of recent vQMC simulations\cite{hu2013}, which directly calculate the spin gaps. Recent DMRG calculations also suggest that there is a gapless spin liquid region\cite{wang2017} in $0.45\lesssim J_2 \lesssim 0.52$. The major difference is that DMRG calculations suggest that there is another VBS state between $0.5\lesssim J_2\lesssim 0.61$ \cite{gong2014} with spin ${\bf S}$=1 gap, which is absent in our calculations.
We note that a recent iPEPS study with U(1) symmetry shows that the intermediate phase is a CVB\cite{haghshenas2017}, while our results based on finite square lattices show the horizontal and vertical directions are isotropic and there is no CVB order.

To summarize, we investigate the phase diagram of spin-1/2 $J_1$-$J_2$ model on square lattice using finite PEPS methods. The recent developed stochastic gradient method allows us to obtain high accurate ground state energies and wave functions. The absence of spin and dimer orders together with power law decay of correlation functions present strong evidences that the intermediate nonmagnetic phase is a gapless spin liquid. However, since the correlation functions
have large correlation lengths in the nonmagnetic region,
we cannot totally exclude the possibility of the existence of
a very weak VBS order in the nonmagnetic phase,
which may need significantly larger system sizes that beyond our current capability. We hope further developed tensor network methods can access larger systems to reexamine these different scenarios.

We would like to thank L. Wang and S.-S. Gong for helpful discussions, and particularly thank Z.-C. Gu for extensive discussions. L. He was supported by the National Key Research and Development Program of China (Grants No. 2016YFB0201202), and Y. Han was supported by the National Science Foundation of China (Grants No. 11474267) . The numerical calculations have been done on the USTC HPC facilities.

%\bibliography{j1j2}

\end{document}